\documentclass[12pt]{article}
\usepackage{graphicx}
\usepackage{subfigure}
\usepackage{feynmp-auto}
\newcommand{\tr}{\mathrm{Tr}}
\title{Weak gravitational interaction of fermions:\\ quantum viewpoint}
\author{T. Olyaei\thanks{t.olyaei@shirazu.ac.ir}~ and A. Azizi\thanks{azizi@shirazu.ac.ir} \\Physics Department, College of Sciences,
    Shiraz University, Shiraz, Iran}
\date{\today}
\begin{document}
\maketitle
\begin{abstract}
Using the quantum theory of linearized gravity, gravitational
interaction differential cross sections of one fermion by another
fermion, a photon and a scalar particle are calculated in the
fermion rest-frame. Then, according to the obtained results, it is
shown that in the lab frame, the gravitational interaction depends
on the spin of the moving particle and is independent of the spin
of the rest particle. After that, on the dependency of the
gravitational interaction of fermion-photon upon the various
states of photons polarization is discussed.
\end{abstract}
\maketitle


\section{Introduction \label{sec:intro}}

In recent decades, many attempts have been made to unify gravity
with other gauge fields. Generally speaking, it can be said that
the main research in the fundamental physics is in this scope. Due
to the fundamental differences between classical and quantum
nature of a typical theory, the assumption that the Universe can
undergo a combined classical and quantum description is
implausible. Since most of fundamental theories are successfully
formulated in the context of quantum mechanics, one believes that
the fundamental theory of gravity should be reformulated in a
quantum mechanical form. For that reason, the quantum mechanical
expression of general relativity is one of the most important
issues of modern theoretical physics.

Based on the interesting consequences of the LIGO experiment
\cite{r15}, in which confirmed one of the challenging predictions
of Einstein gravity, it is logical to respect Einstein general
relativity in the gravitational interactions. Moreover, since most
of experimental tests of gravity are in the weak field limit, the
pioneer efforts are implemented on the quantizing gravity, by
linearizing the Einstein theory of gravity. However, another
non-perturbative and background independent quantum theory of
gravity is based on the so-called Loop quantum gravity \cite{r5.1,
r6.1} and its covariant approach which is known as the spinfoam
formulation \cite{r7.1}--\cite{r10.1}. In this paper, we focus on
the perturbative linearized method.

One of the systematic methods for the quantization of Einstein
gravitational field in the weak field limit, has been established
in a pioneering work by Gupta \cite{r4}. This theory predicts a
massless spin-two particle (the so-called graviton) as an
intermediate particle of gravitational interaction. In this
regard, one can describe the gravitational interaction of
elementary particles in the same way that employs for
electromagnetic interactions in the quantum electrodynamics.
However, the study of gravitational interactions requires
extensive algebraic calculations.

Following the mentioned method, the Lagrangian that describes the
gravitational interaction of a given field can be obtained by weak
field expansion of the general covariant Lagrangian of this field
around the Minkowski spacetime. Subsequently, the conventional
methods of the quantum field theory for linearized gravity are
used to obtain the graviton vertices. Moreover, following
Donoghue's procedure, one finds that the problem of
renormalizability can be solved by regarding the theory of
general relativity as an effective field theory in the low-energy
regime \cite{Donoghue1,Donoghue2}.

So far, some gravitational interactions of elementary particles
have been investigated using the quantum linearized gravity theory
\cite{r5}--\cite{r16}. But differential cross sections of the
gravitational interaction of a fermion with another fermion,
photon, and a scalar particle in the rest frame of the fermion
particle have not been calculated yet. The purpose of this paper
is to calculate the differential cross sections for this
interactions and compare their results with the previously
obtained results for the gravitational interaction of a scalar
particle with another scalar particle, photon, and a fermion in
the rest frame of scalar particle. Thereafter, dependency of
differential cross section on the spin of both moving and rest
particles in gravitational interactions will be discussed. In the
next section, we briefly discuss how to extract the Feynman rules
for graviton vertices. After that, by using these rules, the
mentioned interactions is studied with details. The discussion on
the results of the differential cross sections has also been made
in the last section of the paper. Hereafter, our conventions are:
$\eta_{\mu\nu}=(+, -, -, -)$, and $\hbar=c=1$.


\section{Interaction Lagrangian \label{II}}

In the weak gravitational field regime, the line element of the
Riemann spacetime ($g_{\mu\nu}$) is defined as the Minkowski
metric tensor $\eta_{\mu\nu}$ accompanying with a symmetric
perturbation tensor $h_{\mu\nu}$ as follows \cite{r4}
\begin{equation}\label{equa-4.1}
g_{\mu\nu}=\eta_{\mu\nu} + \kappa h_{\mu\nu},
\end{equation}
where $\kappa^2=32\pi G_N$, and $G_N$ is the Newton gravitational
constant. In this regime and following the Gupta procedure, one is
able to quantize the gravitational field by applying the harmonic
gauge defined as \cite{r4}
\begin{equation}
\partial_{\lambda} h^{\lambda}_{\:\mu}-\frac{1}{2}\partial_{\mu} h^{\lambda}_{\:\lambda}=0.
\end{equation}

The gravitational interaction with a given gauge field is
described by the coupling energy-momentum tensor of that field
with the gravitational field, and is expressed by the following
interaction Lagrangian
\begin{equation}
\mathcal{L}_{int}=-\frac{k}{2}h_{\mu\nu}T^{\mu\nu}.
\end{equation}

The interaction Lagrangian of gravitational field with an
arbitrary gauge field is obtained by the expansion of the general
covariant Lagrangian of that field by using Eq. (\ref{equa-4.1}).
The gravitational interaction Lagrangian for massive scalar
particles is expressed as
\begin{equation}
\mathcal{L}_{int}=\frac{1}{2}h\left[{(\partial^{\mu}\phi)}^{\ast}
(\partial_{\mu}\phi)-m^{2}({\phi}^{\ast}\phi
)\right]-h_{\mu\nu}{(\partial^{\mu}\phi)}^{\ast}(\partial^{\nu}\phi),
\end{equation}
where $ h=h^{\mu}_{\mu} $ and $\phi^{\ast}$ is the complex
conjugate of $\phi$. The same Lagrangian for photons is expressed
as \cite{r12}
\begin{equation}
\mathcal{L}_{int}=-\frac{1}{8}hF^{\mu\nu}F_{\mu\nu}+\frac{1}{2}h^{\tau}_{\nu}F^{\mu\nu}F_{\mu\tau},
\end{equation}
where the Faraday anti-symmetric tensor
$F_{\mu\nu}=\partial_{\mu}A_{\nu}-
\partial_{\nu}A_{\mu} $. In the case of massive spinor fields, in
order to obtain the gravitational interaction Lagrangian, the
tetrad formalism \cite{r3} is also used, the result for this
Lagrangian is given by \cite{r10}
\begin{equation}
\mathcal{L}_{int}=\frac{1}{4}h\left[i(\overline{\Psi }\gamma^{\mu}
\partial _{\mu}\Psi-\partial_{\mu}\overline{\Psi }\gamma^{\mu}\Psi
)-2m\overline{\Psi
}\Psi\right]-\frac{i}{4}h_{\mu\nu}(\overline{\Psi
}\gamma^{\mu}\partial ^{\nu}\Psi-\partial^{\nu}\overline{\Psi
}\gamma^{\mu}\Psi),
\end{equation}
where $ \gamma_{\mu} $ are the ordinary Dirac matrices. The
Feynman rules for graviton vertices that are needed in this paper,
can be derived From this interaction Lagrangians. The related
Feynman rules are listed in the appendix.


\section{Gravitational interaction of fermions}

Here, we investigate the gravitational interactions of non-similar
fermions, fermion-photon and fermion-scalar particle in the
fermion rest frame. In this regard, we use the Feynman rules for
graviton propagator and vertices that are presented in the
appendix.

\begin{figure}
	\newcommand{\marrow}[5]{%
		\fmfcmd{style_def marrow#1 expr p = drawarrow subpath (1/4, 3/4) of p shifted 6 #2 withpen pencircle scaled 0.4;
			label.#3(btex #4 etex, point 0.5 of p shifted 6 #2); enddef;}
		\fmf{marrow#1,tension=0}{#5}}
    \begin{center}
        \subfigure[]{\includegraphics[scale=1]{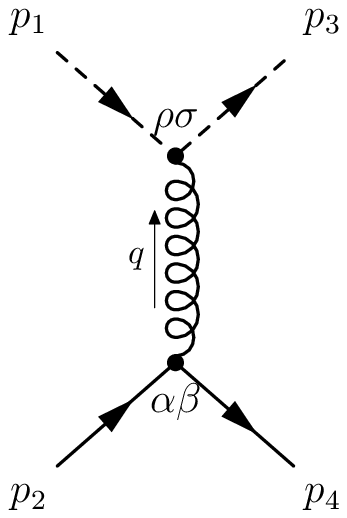}\label{fig-5.2a}}
        \hspace*{2cm}
        \subfigure[]{\includegraphics[scale=1]{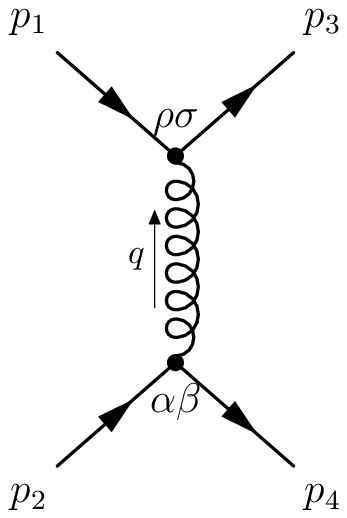}\label{fig-5.2b}}
        \hspace*{2cm}
    \subfigure[]{\includegraphics[scale=1]{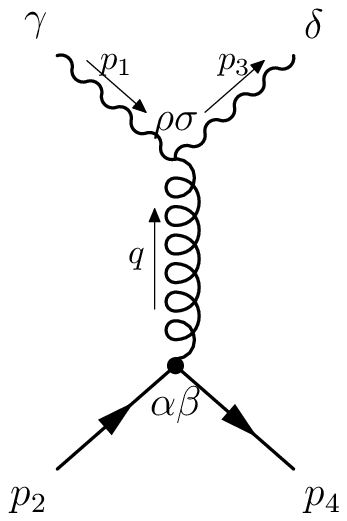}\label{fig-5.2c}}
        \hspace*{2cm}
        \caption{Feynman diagrams for gravitational interactions of (a) fermion-scalar particle, (b) two different fermion particles, (c) fermion-photon.}\label{fig-5.2}
    \end{center}
\end{figure}

First of all, the gravitational interaction of a fermion particle
with mass $ m $ and a scalar particle with mass $ m^{\prime} $ is
considered. The Feynman diagram for this interaction is shown in
Fig. \ref{fig-5.2a}. According to this diagram, the matrix element
for this interaction is written as
\begin{eqnarray}\label{equa-5.44}
    i\mathcal{M}&=&\tau^{\alpha\beta}_{1}(p_{1},p_{3},m^{\prime})\frac{i\mathcal{P}_{\rho\sigma\alpha\beta}}{q^{2}}\bar{u}(p_{4})\tau^{\rho\sigma}_{2}(p_{2},p_{4},m)u(p_{2})\nonumber\\
    &=&\frac{-ik^{2}}{32q^{2}}\bar{u}(p_{4})[-32{m^{\prime}}^{2}m+12{m^{\prime}}^{2}(/\!\!\!p_{2}+/\!\!\!p_{4})+16m(p_{1}\cdot p_{3})\nonumber\\
    &&+4(/\!\!\!p_{3}p_{1}+/\!\!\!p_{1}p_{3})\cdot(p_{2}+p_{4})-8(p_{1}\cdot p_{3})(/\!\!\!p_{2}+/\!\!\!p_{4})]u(p_{2}),
\end{eqnarray}
where $ q=p_{3}-p_{1}=p_{2}-p_{4} $ and the initial and final
four-momenta of the scalar particle are denoted by $ p_{1} $ and $
p_{3} $ and those for the fermion particle are indicated by $
p_{2} $ and $ p_{4} $.

We can use the following equation to obtain the differential cross
section in the rest frame of fermion particle \cite{r17}
\begin{equation}\label{equa-5.76}
\frac{d\sigma}{d\Omega}=\frac{1}{64\pi^{2}m^{2}}\frac{E_{3}^{2}}{E_{1}^{2}}{\overline{|\mathcal{M}|^{2}}},
\end{equation}
where $ E_{1}$ and $ E_{3} $ are the initial and final energies of
the scalar particle and
\begin{equation}\label{equa-5.83}
{\overline{|\mathcal{M}|^{2}}}=\frac{1}{1+2s_{f}}\sum_{f\mathrm{spins}}{|\mathcal{M}|^{2}},
\end{equation}
in which $s_{f}$ indicates the spin of fermion. Using completeness
relation
\begin{equation}\label{equa-5.8}
\sum_{\mathrm{spins}}{u}^{(s)}(p)\bar{u}^{(s)}(p)=/\!\!\!p+m,
\end{equation}
with the following equation
\begin{eqnarray}\label{equa-5.9}
\sum_{{s'}}\bar{u}^{({s'})}_{\alpha}({p'})\gamma^{\mu}_
{\alpha\beta}\sum_{s}u^{(s)}_{\beta}(p)\bar{u}^{(s)}_
{\gamma}(p)\gamma^{\nu}_{\gamma\delta}u^{({s'})}_{\delta}
({p'})
&=&\underbrace{\sum_{{s'}}u^{({s'})}_{\delta}({p'})\bar{u}^{({s'})} _{\alpha}({p'})}_{(/\!\!\!{p'}+{m'})_{\delta\alpha}}\gamma^{\mu}_{\alpha\beta} \underbrace{\sum_{s}u^{(s)}_{\beta}(p)\bar{u}^{(s)}_{\gamma}(p)}_{(/\!\!\!p+m)_{\beta\gamma}}\gamma^{\nu}_{\gamma\delta} \nonumber \\
&=&\tr[(/\!\!\!{p'}+{m'})\gamma^{\mu}(/\!\!\!p+m)\gamma^{\nu}],
\end{eqnarray}
the following expression is obtained for $
{\overline{|\mathcal{M}|^{2}}} $ in fermion-scalar particle interaction
\begin{eqnarray}\label{equa-5.46}
\overline{|\mathcal{M}|^{2}}&=&\frac{2k^{4}}{(64q^{2})^{2}}\tr\{(/\!\!\!p_{4}+m)[16{m^{\prime}}^{2}(/\!\!\!p_{2}+/\!\!\!p_{4}-2m)+16m(p_{1}\cdot p_{3})-4{m^{\prime}}^{2}(/\!\!\!p_{2}+/\!\!\!p_{4})\nonumber\\
&&+4p_{1}\cdot (p_{2}+p_{4})/\!\!\!p_{3}+4p_{3}\cdot(p_{2}+p_{4})/\!\!\!p_{1}-8(p_{1}\cdot p_{3})(/\!\!\!p_{2}+/\!\!\!p_{4})](/\!\!\!p_{2}+m)\nonumber\\
&&\times[16{m^{\prime}}^{2}(/\!\!\!p_{2}+/\!\!\!p_{4}-2m)+16m(p_{1}\cdot p_{3})-4{m^{\prime}}^{2}(/\!\!\!p_{2}+/\!\!\!p_{4})+4p_{1}\cdot(p_{2}+p_{4})/\!\!\!p_{3}\nonumber\\
&&+4p_{3}\cdot(p_{2}+p_{4})/\!\!\!p_{1}-8(p_{1}\cdot p_{3})(/\!\!\!p_{2}+/\!\!\!p_{4})]\}.
\end{eqnarray}
Using \texttt{Tracer} package \cite{r18} in \texttt{Mathematica}
environment, we can compute the traces.

In the fermion rest frame and assuming that the mass of the scalar
particle is negligible compared to the fermion mass, we can write
 \begin{eqnarray}
 &&p_{1}\cdot p_{4}=p_{1}\cdot(p_{1}+p_{2}-p_{3})\longrightarrow p_{1}\cdot p_{4}\simeq p_{1}\cdot p_{2}-p_{1}\cdot p_{3},\label{equa-5.50}\\
 &&p_{2}\cdot p_{4}=p_{2}\cdot(p_{1}+p_{2}-p_{3})\longrightarrow p_{2}\cdot p_{4}=p_{2}\cdot p_{1}+m^{2}-p_{2}\cdot p_{3},\label{equa-5.51}\\
 &&p_{3}\cdot p_{4}=p_{3}\cdot(p_{1}+p_{2}-p_{3})\longrightarrow p_{3}\cdot p_{4}\simeq p_{3}\cdot p_{1}+p_{3}\cdot p_{2},\label{equa-5.52}\\
 &&q^{2}=(p_{1}-p_{3})^{2}=p_{1}^{2}+p_{3}^{2}-2p_{1}\cdot p_{3}\longrightarrow q^{2}\simeq-2p_{1}\cdot p_{3},\label{equa-5.56}\\
 &&p_{1}\cdot p_{2}=mE_{1},\label{equa-5.53}\\
 &&p_{2}\cdot p_{3}=mE_{3},\label{equa-5.54}\\
 &&p_{1}\cdot p_{3}=E_{1}E_{3}-\vec{p_{1}}\cdot\vec{p_{3}}\simeq E_{1}E_{3}-E_{1}E_{3}\cos\theta=2E_{1}E_{3}\sin^{2}\frac{\theta}{2},\label{equa-5.55.1}\\
 &&q+p_{2}=p_{4}\longrightarrow q^{2}=-2p_{2}\cdot q=-2(E_{1}-E_{3})m\label{equa-5.55}.
 \end{eqnarray}
Using Eqs. (\ref{equa-5.56}), (\ref{equa-5.55.1}) and (\ref{equa-5.55}), we get
\begin{equation}\label{equa-5.58}
E_{1}-E_{3}=\frac{2E_{1}E_{3}\sin^{2}\frac{\theta}{2}}{m}.
\end{equation}
After trace completion and inserting Eqs.
(\ref{equa-5.50})--(\ref{equa-5.55.1}) in Eq. (\ref{equa-5.46}),
we obtain
\begin{eqnarray}\label{equa-5.56.2}
\overline{|\mathcal{M}|^{2}}&=&\frac{k^{4}}{64}\left[2mE_{1}E_{3}\sin^{2}\frac{\theta}{2}(E_{3}-E_{1})+m^{2}(E_{1}^{2}+E_{3}^{2}-10E_{1}E_{3})+\frac{6m^{3}}{\sin^{2}\frac{\theta}{2}}(E_{1}-E_{3})\right.\nonumber\\
&&\left.+\frac{4m^{4}}{\sin^{4}\frac{\theta}{2}}\right].
\end{eqnarray}
Using Eq. (\ref{equa-5.58}), we can simplify Eq.
(\ref{equa-5.56.2}) as
\begin{equation}\label{equa-5.59}
\overline{|\mathcal{M}|^{2}}=\frac{k^{4}m^{2}}{16}\left[\frac{(E_{1}-E_{3})m}{2\sin^{2}\frac{\theta}{2}}+\frac{m^{2}}{\sin^{4}\frac{\theta}{2}}\right].
\end{equation}

Assuming that the collision is elastic, the initial and final
energies of the scalar particle are the same. With this condition,
and inserting Eq. (\ref{equa-5.59}) in Eq. (\ref{equa-5.76}), one
can obtain the differential cross section for the gravitational
interaction of fermion-scalar particle as
\begin{equation}\label{equa-5.60}
\frac{d\sigma}{d\Omega}=\frac{k^{4}m^{2}}{(32\pi)^{2}}\frac{1}{\sin^{4}\frac{\theta}{2}}.
\end{equation}
According this relation, one finds decreasing the scattering angle
leads to increasing the differential cross section.

Now, let us proceed our paper with the consideration of the
gravitational interaction of two different fermions with masses $
m $ and $ m^{\prime} $. Considering Fig.~\ref{fig-5.2b}, we obtain
the following expression for matrix element of the interaction
\begin{eqnarray}\label{equa-5.87}
    i\mathcal{M}&=&\bar{u}(p_{3})\tau^{\alpha\beta}_{2}(p_{1},p_{3},m^{\prime})u(p_{1})\frac{i\mathcal{P}_{\rho\sigma\alpha\beta}}{q^{2}}\bar{u}(p_{4})\tau^{\rho\sigma}_{2}(p_{2},p_{4},m)u(p_{2})\nonumber\\
    &=&\frac{-ik^{2}}{64q^{2}}\bar{u}(p_{3}) [2\eta^{\alpha\beta}(/\!\!\!p_{1}+/\!\!\!p_{3}-2m^{\prime})-(p_{1}+p_{3})^{\alpha}\gamma^{\beta}-(p_{1}+p_{3})^{\beta}\gamma^{\alpha}]u(p_{1})\bar{u}(p_{4})\nonumber\\
    &&\times[-\eta_{\alpha\beta}(/\!\!\!p_{2}+/\!\!\!p_{4}-4m)-(p_{2}+p_{4})_{\alpha}\gamma_{\beta}-(p_{2}+p_{4})_{\beta}\gamma_{\alpha}]u(p_{2}),
\end{eqnarray}
where $ p_{1} $ and $ p_{3} $ are the initial and final
four-momenta of the fermion particle with mass $ m^{\prime} $, $
p_{2} $ and $ p_{4} $ are those for the fermion particle with mass
$ m $ and $ q=p_{3}-p_{1}=p_{2}-p_{4} $.

If we investigate this interaction in the rest frame of fermion
particle with mass $ m $, the differential cross section can be
obtained from Eq. (\ref{equa-5.76}), and $
\overline{|\mathcal{M}|^{2}} $ is given by
\begin{equation}\label{equa-5.87.1}
    \overline{|\mathcal{M}|^{2}}=\frac{1}{1+2s_{f}}\frac{1}{1+2s_{f^{\prime}}} \sum_{f\mathrm{spins}}\sum_{f^{\prime}\mathrm{spins}}|\mathcal{M}|^{2}.
\end{equation}
Using Eqs. (\ref{equa-5.8}) and (\ref{equa-5.9}), we can rewrite
Eq. (\ref{equa-5.87.1}) as
\begin{equation}\label{equa-5.87.2}
    \overline{|\mathcal{M}|^{2}}=\frac{k^{4}}{4(64q^{2})^{2}} A^{\alpha\beta\rho\sigma}B_{{\alpha\beta\rho\sigma}},
\end{equation}
where
\begin{eqnarray}\label{equa-5.87.3}
    A^{\alpha\beta\rho\sigma}&=&\tr\{[2\eta^{\alpha\beta}(/\!\!\!p_{1}+/\!\!\!p_{3}-2m^{\prime}) -(p_{1}+p_{3})^{\alpha}\gamma^{\beta}-\gamma^{\alpha}(p_{1}+p_{3})^{\beta}](/\!\!\!p_{1}+m^{\prime})\nonumber\\
    &&\times[2\eta^{\rho\sigma}(/\!\!\!p_{1}+/\!\!\!p_{3}-2m^{\prime})-(p_{1}+p_{3})^{\rho}\gamma ^{\sigma}-\gamma^{\rho}(p_{2}+p_{4})^{\sigma}](/\!\!\!p_{3}+m^{\prime})\},
\end{eqnarray}
\begin{eqnarray}\label{equa-5.87.4}
    B_{\alpha\beta\rho\sigma}&=&\tr\{[-\eta_{\alpha\beta}(/\!\!\!p_{2}+/\!\!\!p_{4}-4m)-(p_{2}+p_{4})_{\alpha}\gamma_{\beta}-\gamma_{\alpha}(p_{2}+p_{4})_{\beta}](/\!\!\!p_{2}+m)\nonumber\\
    &&\times[-\eta_{\rho\sigma}(/\!\!\!p_{2}+/\!\!\!p_{4}-4m)-(p_{2}+p_{4})_{\rho}\gamma_{\sigma}-\gamma_{\rho}(p_{2}+p_{4})_{\sigma}](/\!\!\!p_{4}+m)\}.
\end{eqnarray}

Using the approximation that the rest fermion is so heavy, the
mass of another fermion in the interaction can be neglected and
Eqs. (\ref{equa-5.50})--(\ref{equa-5.58}) are satisfied in this
interaction. By calculating Eqs. (\ref{equa-5.87.3}) and
(\ref{equa-5.87.4}) by using \texttt{Tracer} package and inserting
Eqs. (\ref{equa-5.50})--(\ref{equa-5.55.1}) and (\ref{equa-5.58})
in the obtained expression for Eq. (\ref{equa-5.87.2}), we obtain
\begin{eqnarray}\label{equa-5.87.6}
\overline{|\mathcal{M}|^{2}}&=&\frac{2k^{4}}{(32)^{2}}\left[4m^{2}(E_{2}-E_{4})^2+4m^{3}(E_{2}-E_{4})+\frac{4m^{3}}{\sin^{2}\frac{\theta}{2}}(E_{2}-E_{4})\right.\\\nonumber
&&\left.+\frac{8m^{4}\cos^{2}\frac{\theta}{2}}{\sin^{4}\frac{\theta}{2}}\frac{(E_{2}+E_{4})^2}{E_{2}E_{4}}\right].
\end{eqnarray}
With the assumption that the collision is elastic and inserting
Eq. (\ref{equa-5.87.6}) in Eq. (\ref{equa-5.76}), the differential
cross section for gravitational interaction of non-similar
fermions is obtained as
\begin{equation}\label{equa-5.87.8}
\frac{d\sigma}{d\Omega}=\frac{k^{4}m^{2}}{(32\pi)^{2}}\frac{\cos^{2}\frac{\theta}{2}}{\sin^{4}\frac{\theta}{2}}.
\end{equation}

It is clear that when $\theta$ goes to zero, the differential cross
section diverges. In addition, for the case $\theta=\pi$, unlike
Eq. (\ref{equa-5.60}), the cross section for fermion-fermion
gravitational interaction vanishes.

Following the earlier treatment, we consider the fermion-photon
gravitational interaction where its Feynman diagram is shown in
Fig. \ref{fig-5.2c}. We find that the matrix element of this
diagram is
\begin{eqnarray}\label{equa-5.82}
i\mathcal{M}&=&\bar{u}(p_{4})\tau_{2}^{\rho\sigma}(p_{2},p_{4})u(p_{2})\frac{i\mathcal{P}_{\rho\sigma\alpha\beta}}{q^{2}}\epsilon_{1\gamma}\tau_{3}^{\alpha\beta(\gamma\delta)}(p_{1},p_{3})\epsilon_{3\delta}^{\ast}\nonumber\\
&=&\frac{-ik^{2}}{16q^{2}}\bar{u}(p_{4})[2\eta^{\rho\sigma}(/\!\!\!p_{2}+/\!\!\!p_{4}-2m)-(p_{2}+p_{4})^{\rho}\gamma^{\sigma}-\gamma^{\rho}(p_{2}+p_{4})^{\sigma}]u(p_{2})\nonumber\\
&&\times[p_{1}\cdot p_{3}(\epsilon_{1\sigma}\epsilon_{3\rho}+\epsilon_{1\rho}\epsilon_{3\sigma})-\epsilon_{1}\cdot p_{3}(\epsilon_{3\sigma}p_{1\rho}+\epsilon_{3\rho}p_{1\sigma})-\epsilon_{3}\cdot p_{1}(\epsilon_{1\sigma}p_{3\rho}+\epsilon_{1\rho}p_{3\sigma})\nonumber\\
&&+\epsilon_{1}\cdot\epsilon_{3}(p_{1\sigma}p_{3\rho}+p_{1\rho}p_{3\sigma})+\epsilon_{1}\cdot p_{3}\epsilon_{3}\cdot p_{1}\eta_{\rho\sigma}-\epsilon_{1}\cdot \epsilon_{3}p_{1}\cdot p_{3}\eta_{\rho\sigma}],
\end{eqnarray}
where $ q=p_{3}-p_{1}=p_{2}-p_{4} $, $ p_{1} $ and $ p_{3} $ are
the initial and final propagation four-vectors of photon and $
p_{2} $ and $ p_{4} $ are those for fermion particle, $
\epsilon_{1\gamma} $ and $ \epsilon_{3\delta} $ are the
polarization vectors of photon and $ m $ is the fermion mass.

We study this interaction in the rest frame of the fermion and
regard the following definition for $ p_{1} $, $ p_{2} $ and $
p_{3}$
\begin{eqnarray}
&&p_{1}=(E_{1},0,0,E_{1}),\label{equa-5.64}\\
&&p_{2}=(m,0,0,0),\label{equa-5.65}\\
&&p_{3}=(E_{3},E_{3}\sin\theta,0,E_{3}\cos\theta),\label{equa-5.66}
\end{eqnarray}
and also the circular polarization vectors of photon which are
defined as
\begin{eqnarray}
&&\epsilon_{1+}=\frac{-1}{\sqrt{2}}(0,1,i,0),\label{equa-5.67}\\
&&\epsilon_{1-}=\frac{1}{\sqrt{2}}(0,1,-i,0),\label{equa-5.68}\\
&&\epsilon_{3+}=\frac{1}{\sqrt{2}}(0,-\cos\theta,-i,\sin\theta),\label{equa-5.69}\\
&&\epsilon_{3-}=\frac{1}{\sqrt{2}}(0,\cos\theta,-i,-\sin\theta).\label{equa-5.70}
\end{eqnarray}

Taking into account the recent equations, one finds that there are
four different combinations of polarization vectors of photon for
matrix element.

We can calculate the differential cross section with use of Eq.
(\ref{equa-5.76}), where $ {\overline{|\mathcal{M}|^{2}}} $ is
defined with Eq. (\ref{equa-5.83}) . If the polarized photons
participate in the interaction, the following expression is
obtained for $ {\overline{|\mathcal{M}|^{2}}} $
\begin{equation}\label{equa-5.84}
\overline{|\mathcal{M}|^{2}}=\frac{k^{4}}{2(16q^{2})^{2}}A^{\rho\sigma\mu\nu}B_{{\rho\sigma\mu\nu}},
\end{equation}
where
\begin{eqnarray}\label{equa-5.85}
A^{\rho\sigma\mu\nu}&=&\tr\{[2\eta^{\rho\sigma}(/\!\!\!p_{2}+/\!\!\!p_{4}-2m)- (p_{2}+p_{4})^{\rho}\gamma^{\sigma}-\gamma^{\rho}(p_{2}+p_{4})^{\sigma}](/\!\!\!p_{2}+m)\nonumber\\
&&\times[2\eta^{\mu\nu}(/\!\!\!p_{2}+/\!\!\!p_{4}-2m)-(p_{2}+p_{4})^{\mu}\gamma^{\nu}-\gamma^{\mu}(p_{2}+p_{4})^{\nu}](/\!\!\!p_{4}+m)\},
\end{eqnarray}
\begin{eqnarray}\label{equa-5.86}
B_{\rho\sigma\mu\nu}&=&[p_{1}\cdot p_{3}(\epsilon_{1\sigma}\epsilon_{3\rho}+\epsilon_{1\rho}\epsilon_{3\sigma})-\epsilon_{1}\cdot p_{3}(\epsilon_{3\sigma}p_{1\rho}+\epsilon_{3\rho}p_{1\sigma})-\epsilon_{3}\cdot p_{1}(\epsilon_{1\sigma}p_{3\rho}+\epsilon_{1\rho}p_{3\sigma})\nonumber\\
&&+\epsilon_{1}\cdot \epsilon_{3}(p_{1\sigma}p_{3\rho}+p_{1\rho}p_{3\sigma})+\epsilon_{1}\cdot p_{3}\epsilon_{3}\cdot p_{1}\eta_{\rho\sigma}-\epsilon_{1}\cdot \epsilon_{3}p_{1}\cdot p_{3}\eta_{\rho\sigma}]\nonumber\\
&&\times[p_{1}\cdot p_{3}(\epsilon_{1\nu}\epsilon_{3\mu}+\epsilon_{1\mu}\epsilon_{3\nu})-\epsilon_{1}\cdot p_{3}(\epsilon_{3\nu}p_{1\mu}+\epsilon_{3\mu}p_{1\nu})-\epsilon_{3}\cdot p_{1}(\epsilon_{1\nu}p_{3\mu}+\epsilon_{1\mu}p_{3\nu})\nonumber\\
&&+\epsilon_{1}\cdot \epsilon_{3}(p_{1\nu}p_{3\mu}+p_{1\mu}p_{3\nu})+\epsilon_{1}\cdot p_{3}\epsilon_{3}\cdot p_{1}\eta_{\mu\nu}-\epsilon_{1}\cdot \epsilon_{3}p_{1}\cdot p_{3}\eta_{\mu\nu}].
\end{eqnarray}

After calculating the trace in Eq. (\ref{equa-5.85}), developing
the products in Eq. (\ref{equa-5.86}) and inserting Eqs.
(\ref{equa-5.64})--(\ref{equa-5.70}) in Eq. (\ref{equa-5.84}), we
can obtain the following expressions for the differential cross
sections of gravitational interaction for the fermion-polarized
photon
\begin{eqnarray}
&&\frac{d\sigma}{d\Omega}_{++}=\frac{d\sigma}{d\Omega}_{--}=\frac{E_{3}^{2}}{E_{1}^{2}}\frac{k^{4}m^{2}}{(32\pi)^{2}}\frac{\cos^{4}\frac{\theta}{2}}{\sin^{4}\frac{\theta}{2}},\label{equa-5.96}\\
&&\frac{d\sigma}{d\Omega}_{+-}=\frac{d\sigma}{d\Omega}_{-+}=0,\label{equa-5.97}
\end{eqnarray}
where we have used the notation
$\frac{d\sigma}{d\Omega}_{\lambda_{1}\lambda_{2}} $, in which $
\lambda_{1}=\pm $ means $ \epsilon_{1\pm} $ and $ \lambda_{2}=\pm
$ means $ \epsilon_{2\pm} $.

Eventually, the differential cross section for the gravitational
interaction of fermion-photon with unpolarized photons is given by
\begin{equation}\label{equa-5.100}
\frac{d\sigma}{d\Omega}=\frac{1}{2}\sum_{pol.}\frac{d\sigma}{d\Omega}_{\lambda_{1}\lambda_{3}}=
\frac{k^{4}m^{2}}{(32\pi)^{2}}\frac{\cos^{4}\frac{\theta}{2}}{\sin^{4}\frac{\theta}{2}},
\end{equation}
where we use the assumption that the target is so heavy and its
recoil momentum is neglected. It is clear that the gravitational
interaction of fermion-photon with unpolarized photons is a
quartic function of $ \cos\frac{\theta}{2} $ which is different
comparing to former incoming particles.

\begin{figure}[h]
{\includegraphics[scale=1]{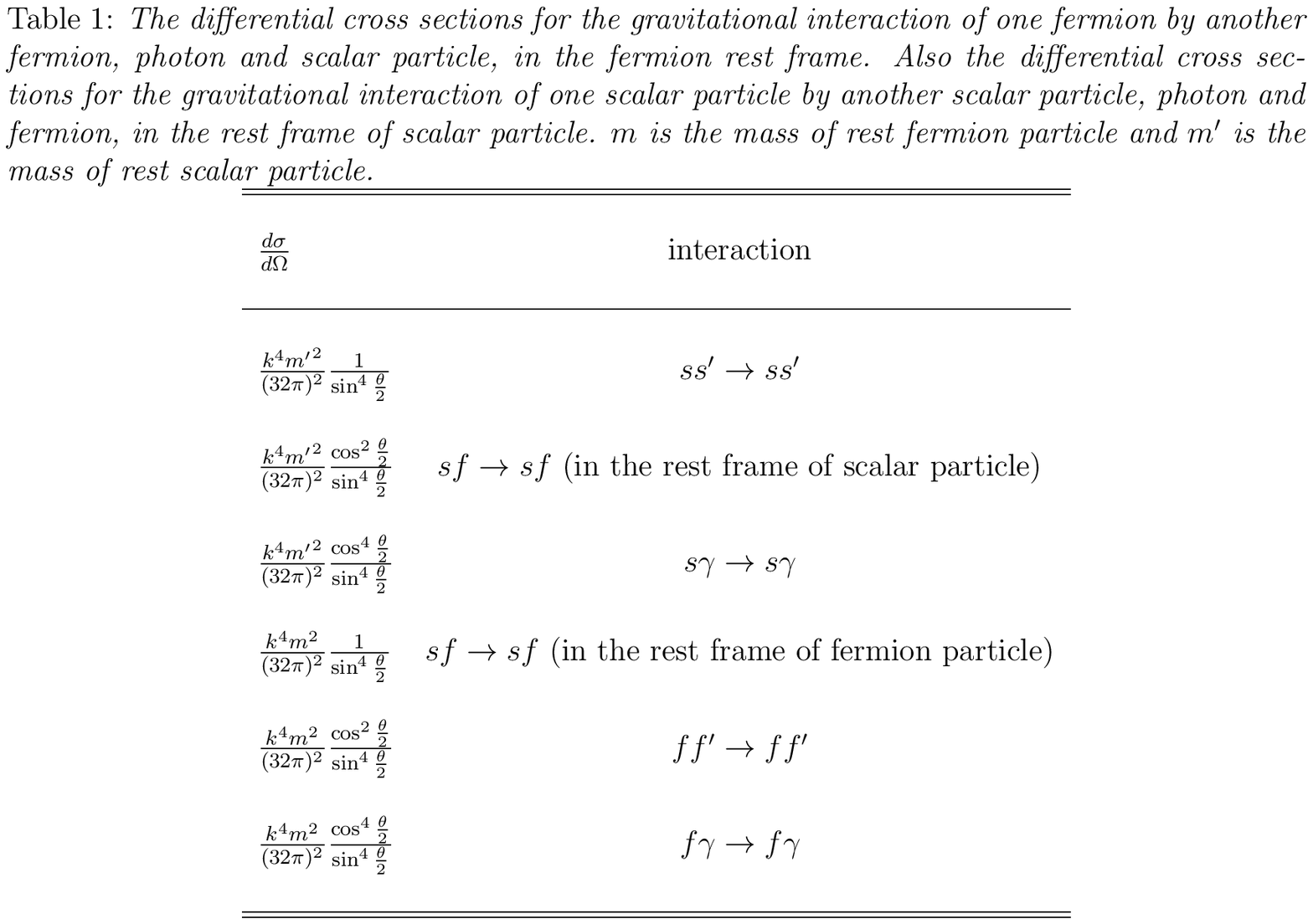}\label{tab1}} 
\end{figure}

The obtained expressions for the cross sections in the present
work are displayed in Table 1. In order to compare these
results with the differential cross sections of gravitational
scattering of different quantum particles by a scalar particle,
the cross sections for these interactions in the approximation of
no recoil of the rest particle is also shown in the Table
1 \cite{r9, r17.1, r18.1}.


\section{Conclusions} \label{IV}

In this paper, we have used the weak field limit of gravitational
interaction by linearized gravity as an effective theory with
quantum gravity motivation. We have obtained the cross sections of
our favorite interactions by using a process that involves
specifying graviton vertices by evaluation of the interaction
Lagrangians of the gravitational field with the other gauge
fields. We have calculated the matrix elements and eventually
obtained the cross sections. These results can now be compared
with the differential cross sections of gravitational scattering
of particles with different spins by a scalar particle; According
to the analytical results in Table 1, it can be concluded
that the differential cross section for the gravitational
interaction of elementary particles is independent of the spin of
the rest particle but depends on the spin of the moving particle.
As a result, the fermion-scalar particle gravitational interaction
in the fermion rest frame and in the scalar particle rest frame
are completely different.

Furthermore, according to Eqs. (\ref{equa-5.96}) and
(\ref{equa-5.97}) we have found that the gravitational interaction
of fermion-photon depends on the polarization of photons, as in
this interaction the polarizations of the incoming and the
outgoing photons are different, the differential cross section
becomes zero, which means that the polarization of the photons in
this interaction remains unchanged.

Moreover, the differential cross sections for small scattering
angle is considered. According to Table 1, the
differential cross sections obtained in this note are large in the
forward scattering. This indicates significant gravitational
effects for very small scattering angle.

As a final remark, we should note that the interactions in the
presented table indicate that the differential cross section for a
gravitational interaction is a function of $
(\cos\frac{\theta}{2})^{4s} $, where $ s $ is the spin of moving
particle in the interaction. In other word, the differential cross
section for a gravitational interaction does not depend on the
spin of target in the target rest frame.

\appendix
\section{Graviton propagator and vertices \label{appb}}

\subsection{Graviton propagator}

\begin{eqnarray}
\parbox{60pt}{\includegraphics[scale=1]{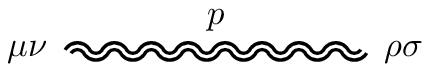}} \qquad\qquad\qquad &=& \frac{i\mathcal{P}_{\mu \nu \rho \sigma}}{p^2+i\epsilon}\nonumber\\
\mathcal{P}_{\mu \nu \rho \sigma} &\equiv& \frac{1}{2}({\eta}_{\mu\rho}{\eta}_{\nu\sigma} + {\eta}_{\mu\sigma}{\eta}_{\nu\rho}-{\eta}_{\mu\nu}{\eta}_{\rho\sigma}). \label{a1}
\end{eqnarray}

\subsection{Scalar-scalar-graviton vertex}

\begin{eqnarray}
\parbox{60pt}{\includegraphics[scale=1]{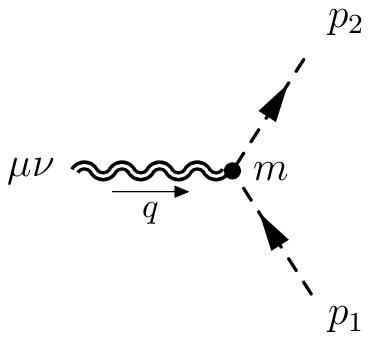}} \qquad &=& \tau^{\mu\nu}_{1}(p_1,p_2,m)\nonumber\\
&=& \frac{i}{2}k[\eta^{\mu\nu}(p_{1}\cdot p_{2}-m^{2})-p_{1}^{\mu}p_{2}^{\nu}-p_{1}^{\nu}p_{2}^{\mu}].
\label{a2}
\end{eqnarray}

\subsection{Fermion-fermion-graviton vertex}

\begin{eqnarray}
\parbox{60pt}{\includegraphics[scale=1]{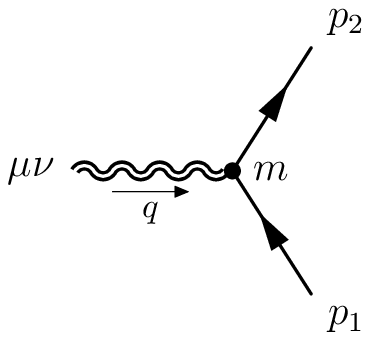}} \qquad &=& \tau^{\mu \nu}_{2}(p_1,p_2,m)\nonumber\\
&=& \frac{i}{8}\kappa \left[2\eta^{\mu\nu}(/\!\!\!p_{1} + /\!\!\!p_{2} - 2m) - (p_{1} + p_{2})^{\mu}\gamma^{\nu} - (p_{1} + p_{2})^{\nu}\gamma^{\mu}\right].
\label{a3}
\end{eqnarray}

\subsection{Photon-photon-graviton vertex}

\begin{eqnarray}
\parbox{60pt}{\includegraphics[scale=1]{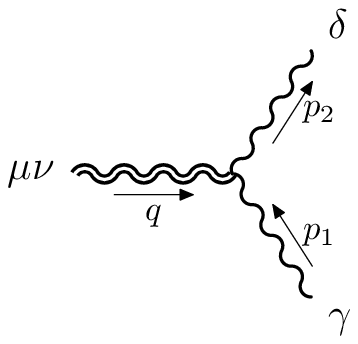}} \qquad\ &=& \tau^{\mu \nu}(p_1,p_2,m)\nonumber\\
&=& \frac{i}{2}k[2\mathcal{P}^{\mu\nu(\gamma\delta)}(p_{1}\cdot p_{2})
+\eta^{\mu\nu}p_{1}^{\delta}p_{2}^{\gamma}+\eta^{\gamma\delta}(p_{1}^{\mu}p_{2}^{\nu}+p_{1}^{\nu}p_{2}^{\mu})\nonumber\\
&&-(\eta^{\mu\delta}p_{1}^{\nu}p_{2}^{\gamma}+\eta^{\nu\delta}p_{1}^{\mu}p_{2}^{\gamma}+\eta^{\nu\gamma}p_{1}^{\delta}p_{2}^{\mu}+\eta^{\mu\gamma}p_{1}^{\delta}p_{2}^{\nu})].
\label{a4}
\end{eqnarray}


%
\end{document}